# Force law in material media and quantum phases

Alexander L Kholmetskii[1], Oleg V. Missevitch[2] and T Yarman[3,4]

[1]Belarusian State University, Minsk, Belarus, e-mail: alkholmetskii@gmail.com
[2]Institute for Nuclear Problems, Belarusian State University, Minsk, Belarus
[3]Okan University, Akfirat, Istanbul, Turkey
[4]Savronik, Eskisehir, Turkey

We show that the known expressions for the force on a point-like dipole are incompatible with the relativistic transformation of force, and in this respect we apply the Lagrangian approach to the derivation of the correct equation for force on a small electric/magnetic dipole. The obtained expression for the generalized momentum of a moving dipole predicts two novel quantum effects with non-topological and non-dynamic phases, when an electric dipole is moving in an electric field, and when a magnetic dipole is moving in a magnetic field, correspondingly. The implications of the obtained results are discussed.

*1. Introduction.* – As is known, the Lorentz force in a material medium is given by the equation
$$f^\alpha = F^\alpha{}_\beta j^\beta \quad (\alpha, \beta = 0\ldots 3), \tag{1}$$
where $F^\alpha{}_\beta$ is the tensor of electromagnetic field, and $j^\beta$ is the four-current density, which in the case of bound charged is determined via the relationship
$$j^\beta = \partial_\alpha M^{\alpha\beta}, \tag{2}$$
and $M^{\alpha\beta}$ is the magnetization-polarization tensor [1, 2].

For an electrically neutral compact bunch of charges with time-independent proper electric $p_0$ and magnetic $m_0$ dipole moments, the spatial components of eq. (1) integrate to the 3d-force [3, 4]
$$\boldsymbol{F} = (\boldsymbol{p}\cdot\nabla)\boldsymbol{E} + \nabla(\boldsymbol{m}\cdot\boldsymbol{B}) + \frac{1}{c}(\boldsymbol{p}\times(\boldsymbol{v}\cdot\nabla)\boldsymbol{B}), \tag{3}$$
where $\boldsymbol{p} = \boldsymbol{p}_0 - \frac{(\gamma-1)}{\gamma v^2}(\boldsymbol{p}_0\cdot\boldsymbol{v})\boldsymbol{v} + \frac{\boldsymbol{v}\times\boldsymbol{m}_0}{c}$, $\boldsymbol{m} = \boldsymbol{m}_0 - \frac{(\gamma-1)}{\gamma v^2}(\boldsymbol{m}_0\cdot\boldsymbol{v})\boldsymbol{v} + \frac{\boldsymbol{p}_0\times\boldsymbol{v}}{c}$ (4a-b)

are the electric and magnetic dipole moments, correspondingly, in an inertial frame of observation, wherein the dipole moves with the velocity $\boldsymbol{v}$, and $\gamma$ is the Lorentz factor.

We point out that eq. (3) does not include the force component due to time variation of hidden momentum of magnetic dipoles, which, however, is strongly required by the general relativistic theorem [5, 6] about zero total momentum of any isolated static configuration of charges and bound currents. A "manual" (according to the terminology of [7]) addition of this force component [6, 8-10] to the rhs of eq. (3), leading to
$$\boldsymbol{F} = (\boldsymbol{p}\cdot\nabla)\boldsymbol{E} + \nabla(\boldsymbol{m}\cdot\boldsymbol{B}) + \frac{1}{c}(\boldsymbol{p}\times(\boldsymbol{v}\cdot\nabla)\boldsymbol{B}) - \frac{1}{c}\left(\boldsymbol{m}\times\frac{d\boldsymbol{E}}{dt}\right), \tag{5}$$
tacitly recognizes that the force (3) does not represent a derivative of *total* momentum on time and thus, it may turn out to be incompatible with the law of total energy-momentum conservation and relativistic transformation of force.

In a consistent way, the addition of hidden momentum contribution to the Lorentz force should be done in eq. (1), with the corresponding four-force density $f_h^\alpha = -dp_h^\alpha/d\tau$, where $\tau$ is the proper time, and $p_h^\alpha$ must compose a four-vector of hidden energy-momentum density. At the same time, to the moment nobody provided an explicit expression for $p_h^\alpha$, and the failure to solve this problem can be explained by the closed link between hidden momentum density and field momentum density $p_f^\alpha$. However, in the case of bound electromagnetic (EM) field, which is implied in the analysis of the problem of hidden momentum, the components of $p_f^\alpha$ do not compose a four-vector [1]. In these conditions, we cannot assert that eq. (5) (which can be named as the expanded Lorentz force (ELF)) is compatible with the relativistic transformation of force.

Addressing now to the Einstein-Laub force law [11, 12], one can show that at the considered case $p_0$, $m_0$=const, the density of this force integrates to [4]

$$\boldsymbol{F}_{EL} = (\boldsymbol{p}\cdot\nabla)\boldsymbol{E} + (\boldsymbol{m}\cdot\nabla)\boldsymbol{B} + \frac{1}{c}\boldsymbol{p}\times(\boldsymbol{v}\cdot\nabla)\boldsymbol{B} - \frac{1}{c}\boldsymbol{m}\times(\boldsymbol{v}\cdot\nabla)\boldsymbol{E}, \tag{6}$$

Using the vector identity $(\boldsymbol{m}\cdot\nabla)\boldsymbol{B} = \nabla(\boldsymbol{m}\cdot\boldsymbol{B}) - \boldsymbol{m}\times(\nabla\times\boldsymbol{B})$, and Maxwell equation $\nabla\times\boldsymbol{B} = \partial\boldsymbol{E}/c\partial t$, which is valid at the location point of dipole, we arrive exactly at eq. (5), which, in return, via the identity $(\boldsymbol{p}\cdot\nabla)\boldsymbol{E} = \nabla(\boldsymbol{p}\cdot\boldsymbol{E}) - \boldsymbol{p}\times(\nabla\times\boldsymbol{E})$ and Maxwell equation $\nabla\times\boldsymbol{E} = -\partial\boldsymbol{B}/c\partial t$ can be presented in a more symmetric form

$$\boldsymbol{F} = \nabla(\boldsymbol{p}\cdot\boldsymbol{E}) + \nabla(\boldsymbol{m}\cdot\boldsymbol{B}) + \frac{1}{c}\frac{d}{dt}(\boldsymbol{p}\times\boldsymbol{B}) - \frac{1}{c}\frac{d}{dt}(\boldsymbol{m}\times\boldsymbol{E}), \tag{7}$$

where the case of time-dependent $p_0$, $m_0$ is also included [13]. Thus, the Einstein-Laub force law and the expanded Lorentz force law both yield eq. (7). Therefore, the introduced abbreviation ELF for the force (5), being also applicable to eq. (7), can be decoded as the Einstein-Laub force, too, if a reader desires to act so.

*2. On relativistic non-adequacy of ELF.* – Here we show that ELF is incompatible with the relativistic transformation of force[1], which yields relativistically non-adequate situations.

Consider the motion of an electric dipole $p_0$ near a homogeneously charged straight line, as shown in Fig. 1. In the frame $K_0$, the line produces the electric field $E_{0y} = 2\lambda_0/y_0$ at the location of dipole, and, according to eq. (7), there is a single non-vanishing force component, acting on the electric dipole along the axis $y$

$$F_{0y} = \frac{\partial}{\partial y}(\boldsymbol{p}\cdot\boldsymbol{E}) = \frac{2\lambda_0 p_0}{y_0^2 \gamma_u}, \tag{8}$$

where $\gamma_u = (1 - u^2/c^2)^{-1/2}$.

Next, we determine the force on the dipole for an observer in the frame $K$, wherein the charged line moves at the constant velocity $v$ along the axis $x$, while the dipole has the velocity $\boldsymbol{V}\{v, u/\gamma_v, 0\}$, where $\gamma_v = (1 - v^2/c^2)^{-1/2}$.

In this frame, the non-vanishing field components at the location of dipole are as follows:
$$E_y = 2\lambda_0\gamma_v/y_0, \quad B_z = 2\lambda_0 v\gamma_v/cy_0 \tag{9), (10}$$

According to eqs. (4a-b), the electric dipole moment $\boldsymbol{p}$ and the magnetic dipole moment $\boldsymbol{m}$ have the components $\{0, -p_0/\gamma_u, 0\}$, $\{0, p_0 v/c\gamma_u, 0\}$; correspondingly. Then, in the frame K, eq. (7) along with eqs. (9)-(10) yields the following $y$- and $x$-components of force on the dipole:

$$F_y = \frac{\partial}{\partial y}(\boldsymbol{p}\cdot\boldsymbol{E}) + \frac{\partial}{\partial y}(\boldsymbol{m}\cdot\boldsymbol{B}) = -\frac{p_0}{\gamma_u}\frac{\partial E_y}{\partial y} + \frac{p_0 v}{\gamma_u c}\frac{\partial B_z}{\partial y} = \frac{2\lambda_0 p_0}{\gamma_v y_0^2 \gamma_u}, \tag{11}$$

$$F_x = -\frac{1}{c}\frac{d}{dt}\left(\frac{p_0 B_z}{\gamma_u}\right) + \frac{1}{c}\frac{d}{dt}\left(\frac{p_0 v E_y}{c\gamma_u}\right) = 0. \tag{12}$$

---

[1] We point out that this transformation is applicable to any particles with non-variable internal structure (e.g., point-like particles) and, in general, it can be violated, when an internal structure varies with time, giving rise to the corresponding variation of the energy-momentum components, lying beyond the approximation of point-like particles. The latter situation can be referred to the case, where $dp_0/dt$, $dm_0/dt \neq 0$ (see section 3). However, at $p_0, m_0 = const$ (i.e., the case just considered in this section), the violation of relativistic transformation of force would indicate the incorrectness of EFL.

Comparing the obtained results (11), (12) with the relativistic transformation of force [1], we find that for the configuration of Fig. 1, this transformation yields $F_y = F_{0y}/\gamma_v$, which agrees with eqs. (8) and (11). However, the x-component of force should be equal to $F_x = F_{0y}uv/c^2$, which is clearly at odds with eq. (12).

The non-adequacy of eq. (12) can be seen from the fact that at the vanishing x-component of force, the x-component of momentum of dipole should be conserved in K, i.e.

$$\frac{d}{dt}(\gamma_V Mv) = 0,$$

where $\gamma_V = (1 - V^2/c^2)^{-1/2}$, and $M$ is the rest mass of the dipole. Hence, due to the constancy of $v$, we derive $d\gamma_V/dt = 0$, which is obviously wrong, because the dipole is accelerated along the y-axis due to the force (11).

Looking closer at the structure of eq. (12), we can find that the compatibility with the relativistic transformation of force is restored, when the second term on its rhs is eliminated, i.e. the force due to the variation of hidden momentum should be excluded. For the problem considered, this force component emerges in the frame K due to the non-vanishing value of magnetic dipole moment of a moving electric dipole.

Further, we can consider the modification of the problem in Fig. 1, where the straight charged line is replaced by a conducting line, carrying a steady current, while the electric dipole $p_0$ is replaced by the magnetic dipole $m_0\{0,0,-m_0\}$.

Omitting the straightforward calculations, similar to the problem considered just above, we find that the y-component of force on the dipole, calculated in the frames $K_0$ and K with eq. (7), agrees with the relativistic transformation of force. However, the x-component of this force, defined as the sum of the third and fourth terms of eq. (7), is again equal to zero in the frame K. The compatibility with the relativistic transformation of force is restored, when the third term of eq. (7), emerging due to the electric dipole moment of a moving magnetic dipole, is eliminated.

Thus, the relativistic non-adequacy of ELF (7) becomes apparent, which indicates the failure of direct extension of the Lorentz force law for free charges (where it is definitely correct) to a magnetized/polarized medium. In this respect, we apply the most general way to the derivation of force law in material media, based on a postulated Lagrangian.

*3. Force law in a magnetized/polarized medium via the Lagrangian approach.* – We use the Lagrangian density $l$ for an electrically neutral material medium with bound charges in the form (see e.g. [14])

$$l = l_{free} + \frac{1}{2} M^{\alpha\beta} F_{\alpha\beta}, \tag{13}$$

where $l_{free}$ is the Lagrangian density in the absence of external fields. For a compact bunch of charges, having the proper electric $p_0$ and magnetic $m_0$ dipole moments, and moving in an EM field with velocity $v$, eq. (13) integrates to the Lagrangian

$$L = -\frac{Mc^2}{\gamma} + p \cdot E + m \cdot B. \tag{14}$$

where $M$ is the rest mass of the dipole, and $p$, $m$ are determined by eqs. (4a-b).

We determine the force via the equation

$$\frac{\partial L}{\partial r} = \frac{d}{dt}\frac{\partial L}{\partial v}, \tag{15}$$

where we obtain

$$\frac{\partial L}{\partial r} = \nabla(p \cdot E) + \nabla(m \cdot B), \tag{16}$$

$$\frac{\partial L}{\partial v} = \gamma Mv - \frac{\gamma(p_{0//} \cdot E)v}{c^2} - \frac{\gamma(m_{0//} \cdot B)v}{c^2} + \frac{1}{c}(m_0 \times E) - \frac{1}{c}(p_0 \times B). \tag{17}$$

Deriving the latter equation, we presented the electric and magnetic dipole moments in the form

$$p = \frac{p_{0/\!/}}{\gamma} + p_{0\perp} + \frac{v \times m_0}{c}, \quad m = \frac{m_{0/\!/}}{\gamma} + m_{0\perp} + \frac{p_0 \times v}{c},$$

resulting from eqs. (4a-b), where the subscripts "//", "⊥" denote the components, being either collinear, or orthogonal to $v$.

Combining eqs. (15)-(17), we obtain the novel expression for the force on a dipole

$$F = \frac{d}{dt}(\gamma M v) = \nabla(p \cdot E) + \nabla(m \cdot B) + \frac{d}{dt}\frac{\gamma(p_{0/\!/} \cdot E)v}{c^2} + \frac{d}{dt}\frac{\gamma(m_{0/\!/} \cdot B)v}{c^2}$$
$$+ \frac{1}{c}\frac{d}{dt}(p_0 \times B) - \frac{1}{c}\frac{d}{dt}(m_0 \times E),$$
(18)

derived with the invariant Lagrangian (14).

The physical meaning of the force components in eq. (18) can be disclosed via the analysis of the generalized momentum (17), where the last two terms are well understandable[2]. However, in contrast to analogous terms in eq. (7), the electric and magnetic hidden momenta depend only on the proper dipole moments, and do not include the components $v \times m_0/c$ for a moving magnetic dipole and $p_0 \times v/c$ for a moving electric dipole. The absence of these terms in the expressions for electric and magnetic hidden momenta provides the relativistically consistent solution of both problems presented in section 2.

Next contribution to the generalized momentum is supplied by the second and third terms of eq. (17), and the best way to understand their physical meaning is to analyze the problems, where these terms solely determine the total force on the dipole.

Consider, for example, a magnetic dipole $m_0$ with the proper mass $M$, which is oriented along the constant magnetic field $B_0$, and both these vectors are parallel to the axis $x$ of the rest frame of the dipole $K_0$. In this frame, no force and no torque is exerted on the dipole, and its total energy in the magnetic field is equal to
$U_0 = Mc^2 - (m_0 \cdot B_0)$.

In another inertial frame K, wherein the dipole is moving with the constant velocity $v$ along the axis $x$, its total energy is $U = \gamma U_0$, and total momentum

$$P_x = \gamma U_0 v/c^2 = \gamma M v - \gamma v(m_0 \cdot B_0)/c^2.$$
(19)

Now we assume that the magnetic dipole represents two small disks with opposite charges, rotating in the opposite directions around a common axis and, due to their friction, the disks slowly decelerate with a negligible radiation [6, 9]. In this process, the conservation of total energy signifies that the rotational energy of disks is transformed into their heating, while the conservation of momentum (19) yields the equality

$$\frac{d}{dt}(\gamma M v) = \frac{d}{dt}(\gamma v(m_0 \cdot B_0)/c^2) = \frac{d}{dt}(\gamma v(m_{0/\!/} \cdot B)/c^2),$$
(20)

which is just eq. (18) for the problem considered in the frame K. (Here $B\{B_x,0,0\}$, and $B_x = B_{0x}$). Further, due to the constancy of $v$ in the frame K, eq. (20) signifies the decrease of the rest mass of the dipole

$$\frac{dM}{dt} = \frac{d}{dt}\frac{(m_{0/\!/} \cdot B)}{c^2},$$

which is related to the corresponding decrease of mechanical stress energy in the decelerated disks, emerging due to the interaction of moving charges of both disks with the magnetic field. (We notice that eq. (20), in general, is not compatible with the relativistic transformation of forces for the reasons mentioned in the footnote 1).

Consider the modification of this problem, where the vectors $m_0$ and $B_0$, being parallel to

---

[2] In ref. [13], we named the product $-(p \times B)/c$ as the "latent momentum" of electric dipole. Now, we find more appropriate to call the product $(m_0 \times E)/c$ as the magnetic hidden momentum, and the product $-(p_0 \times B)/c$ as the electric hidden momentum for the reasons, which we will explain elsewhere.

each other, are both orthogonal to velocity $v$. Then we similarly obtain in the frame K

$$\frac{d}{dt}(\gamma M v) = \frac{d}{dt}\left(\gamma v(\mathbf{m}_{0\perp} \cdot \mathbf{B}_0)/c^2\right). \quad (21)$$

In order to express eq. (21) via the fields in the frame K, we use the identity $\mathbf{a}\times(\mathbf{b}\times\mathbf{c})=\mathbf{b}(\mathbf{a}\cdot\mathbf{c})-\mathbf{c}(\mathbf{a}\cdot\mathbf{b})$, and then the relativistic transformation of fields between $K_0$ and K. Hence, we obtain

$$\frac{d}{dt}(\gamma M v) = -\frac{1}{c}\frac{d}{dt}(\mathbf{m}_0 \times \mathbf{E}),$$

which shows that at $\mathbf{m}_0 \perp \mathbf{v}$, the process of deceleration of disks, composing a magnetic dipole, is related to the force component due to time variation of magnetic hidden momentum, as already was found, e.g., in ref. [6].

In a similar way we can grasp the meaning of the term $\gamma(\mathbf{p}_{0//} \cdot \mathbf{E})\mathbf{v}/c^2$ in eq. (17).

*4. Quantum phases for an electric/magnetic dipole.* - Continuing to analyze equation (17), we notice that the term $(\mathbf{m}_0 \times \mathbf{E})/c$ (magnetic hidden momentum) is related to the Aharonov-Casher (A-C) effect [15], while the term $-(\mathbf{p}_0 \times \mathbf{B})/c$ (electric hidden momentum) gives rise to the He-McKellar-Wilkens (HMW) quantum phase [16, 17], and both of these effects had been observed [18, 19].

At the same time, eq. (17) additionally contains the components $-\gamma(\mathbf{p}_{0//} \cdot \mathbf{E})\mathbf{v}/c^2$, $-\gamma(\mathbf{m}_{0//} \cdot \mathbf{B})\mathbf{v}/c^2$, which thus should be related to the corresponding quantum phases even in the case, where the classical force is vanishing (e.g., at constant fields $\mathbf{E}, \mathbf{B}$).

In particular, through the motion of an electric dipole $\mathbf{p}_0$ in an electric field $\mathbf{E}$, the quantum phase has the form

$$\delta_{pE} = -\frac{\gamma}{\hbar c^2}\int(\mathbf{p}_{0//} \cdot \mathbf{E})\mathbf{v} \cdot d\mathbf{s}, \quad (22)$$

where $d\mathbf{s}$ is the path element. Analogously, through the motion of a magnetic dipole $\mathbf{m}_0$ in a magnetic field $\mathbf{B}$, the corresponding phase reads as

$$\delta_{mB} = -\frac{\gamma}{\hbar c^2}\int(\mathbf{m}_{0//} \cdot \mathbf{B})\mathbf{v} \cdot d\mathbf{s}. \quad (23)$$

At constant $\mathbf{p}_0, \mathbf{m}_0, \mathbf{E}$ and $\mathbf{B}$, and $\mathbf{v}//d\mathbf{s}$, we obtain in the non-relativistic case ($\gamma\approx 1$)

$$\delta_{pE} \approx -\frac{v}{\hbar c^2}p_{0//}E_{//}L = -\frac{v}{\hbar c^2}p_{0//}\Delta\varphi, \quad (24)$$

and

$$\delta_{mB} \approx -\frac{v}{\hbar c^2}m_{0//}B_{//}L, \quad (25)$$

where $L$ is the path length of dipole in the field region, and $\Delta\varphi$ is the potential difference between the initial and end points of the path, lying in the field region.

We see that $\delta_{pE}, \delta_{mB}$ depend on velocity $\mathbf{v}$ of a dipole and hence, unlike to the Aharonov-Bohm phase [20], A-C and HMW phases, they do not represent topological phases. Besides, they cannot be referred to dynamic phases, insofar as the classical force on a dipole can be equal to zero, while the values of $\delta_{pE}$ and $\delta_{mB}$ are not vanishing. In addition, these phases depend on mutual orientations of vectors $\mathbf{p}_0, \mathbf{E}, \mathbf{m}_0, \mathbf{B}$ and $\mathbf{v}$. These observations allow us to assume that eqs. (22), (23) determine a new kind of quantum phases, which, like A-C and HMW phases, are related to each other via electric-magnetic duality transformations $\mathbf{E}\to\mathbf{B}, \mathbf{p}\to\mathbf{m}$ [1].

From the experimental viewpoint, both phases (24) and (25) can be reliably detected.

The phase (24) can be measured via the quantum interference of molecules (e.g., BaS with $p_0\approx 3\times 10^{-29}$ C·m [19]). Then, at $v/c\approx 10^{-4}$, and $\Delta\varphi\approx 10^5$ V, we obtain

$\delta_{pE}\approx 10$ mrad. (26)

The measurement of the phase (25) can be done via the neutron interferometry. For exam-ple, for a neutron beam, passing a solenoid along its axis with the velocity $v \approx 10^{-4}c$, we obtain at $B=10$ T, $L=1$ m and $m_0=9.6\times10^{-24}$ J/T,

$$\delta_{mB} \approx 30 \text{ mrad}. \tag{27}$$

The estimated phases (26), (27) are even larger than typical phases for the A-C and HMW effects. At the same time, the phase $\delta_{pE}$ can be mixed with the Stark phase [21], while the phase $\delta_{pE}$ can be mixed with the Zeeman phase [19]. At the same time, the Stark and Zeeman phases are determined by the general expression

$$\delta \sim \frac{1}{\hbar}\int H dt,$$

where $H$ is the corresponding Hamiltonian and in a real situation, the values of these phases are *inversely proportional* to the velocity of dipole $v$. In contrast, the phases (26), (27) are *directly proportional* to $v$, which provides us with experimental tools (not discussed here for brevity) for separation of these phases.

One more option for the identification of the phases (22) and (23) is related to their sensitivity to the mutual orientation between vectors $p_0$, $E$ and $v$ for the phase (22), and between vectors $m_0$, $B$ and $v$ for the phase (23). This property allows us to obtain additional arguments in the favour of the force law (18) versus the ELF (7). Indeed, eq. (7) implies that the generalized momentum of dipole should contain the sum of the terms

$$\frac{1}{c^2}\left((v \times m_0) \times B - (p_0 \times v) \times E\right), \tag{28}$$

resulting from the substitution of transformations (4a-b) into the third and fourth terms of eq. (7). Using the vector identity $a \times (b \times c) = b(a \cdot c) - c(a \cdot b)$, we derive in the case, where $p_0$, $E \perp v$, and $m_0$, $B \perp v$:

$$\frac{1}{c^2}\left((v \times m_0) \times B - (p_0 \times v) \times E\right) = -\frac{v(m_{0\perp} \cdot B)}{c^2} - \frac{v(p_{0\perp} \cdot E)}{c^2}.$$

We note that at $p_0 // v$, and $m_0 // v$, the contribution to the generalized momentum (28) disappears. Hence, the corresponding phases resulting from eq. (28) read as

$$\delta_{pE}(\text{ELF}) = -\frac{1}{\hbar c^2}\int (p_{0\perp} \cdot E) v \cdot ds, \tag{29}$$

$$\delta_{mB}(\text{ELF}) = -\frac{1}{\hbar c^2}\int (m_{0\perp} \cdot B) v \cdot ds. \tag{30}$$

The comparison of eqs. (22) (defining $\delta_{pE}$) and (29) (defining $\delta_{pE}(\text{ELF})$) shows that through the motion of the electric dipole along the normal of a parallel plate charged capacitor ($p_0$, $E//v$), the phase (22) takes the maximal value, and the phase (29) is equal to zero, whereas at the motion of electric dipole along the axis of a parallel plate charged capacitor ($p_0$, $E \perp v$), the phase (22) is vanishing, while the phase (29) becomes maximal. Thus, this experiment will allow to make a crucial choice between the expressions (18) and (7).

In a similar way, we can compare the magnetic phases (23) and (30), and get one more possibility to test the force law (18) versus ELF (7).

*5. Conclusion.* – We have demonstrated the incompatibility of EFL (7) with the relativistic transformation of force and, in this respect, we applied the most general way to the derivation of the force law in material media based on the Lagrangian density (13). The obtained novel expression for the force on a small dipole (18), and the corresponding expression for the generalized momentum (17) imply the existence of two novel quantum effects with non-topological phases (20) and (23). Comparably large values of the phases $\delta_{pE}$, $\delta_{mB}$, and their specific properties discussed above make possible their experimental detection with the technique of quantum interference of molecules (for $\delta_{pE}$) and, for example, with neutron interferometry (for $\delta_{mB}$).

Finally, one can introduce the common designation for the quantum phases, associated with the motion of dipoles in EM field: $\delta_{pE}$, $\delta_{mB}$, $\delta_{pB}$ (HMW phase), $\delta_{mE}$ (A-C phase), which corres-pond to all possible combinations of the pair *p*, *m* with the pair *E*, *B*, and all of these phases are commonly derived from the generalized momentum (17).

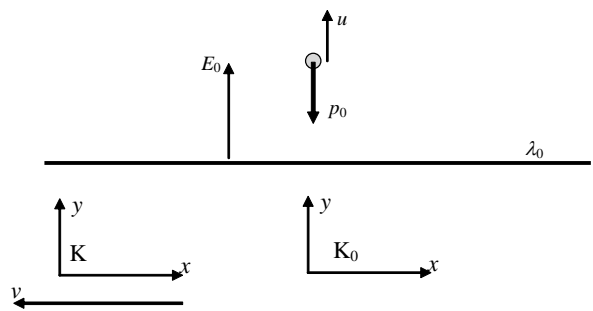

Fig. 1. An electric dipole $\boldsymbol{p}_0\{0,-p_0, 0\}$ is moving with the velocity $\boldsymbol{u}\{0,u, 0\}$ with respect to homogeneously charged long line oriented along the axis $x$, and having the linear charge density $\lambda_0$, as viewed in the inertial frame $K_0$ (the rest frame of the line). We want to calculate the force on the dipole in the frame $K_0$, as well as in another inertial frame K, moving with respect to $K_0$ with constant velocity $v$ in the negative $x$-direction.